\documentclass[twocolumn,showpacs,preprintnumbers,amsmath,amssymb,pre]{revtex4}
\usepackage{graphicx}
\usepackage{dcolumn}
\usepackage{bm}
\usepackage{color}

\begin{document}

\title{Comment on: \textquotedblleft Effect of polydispersity on the ordering transition of adsorbed self-assembled rigid rods\textquotedblright}

\author{L. G. L\'opez}
\email{lglopez@unsl.edu.ar}
\author{D. H. Linares}
\author{A. J. Ramirez-Pastor}
\email{antorami@unsl.edu.ar}
\affiliation{Departamento de
F\'{\i}sica, Instituto de F\'{\i}sica Aplicada, Universidad
Nacional de San Luis, CONICET, 5700 San Luis, Argentina}

\date{\today}

\begin{abstract}

The critical behavior of self-assembled rigid rods on a square
lattice was recently reinvestigated by Almarza et al. [Phys. Rev.
E {\bf 82}, 061117 (2010)]. Based on the Binder cumulants and the
value of the critical exponent of the correlation length, the
authors found that the isotropic-nematic phase transition
occurring in the system is in the two-dimensional Ising
universality class. This conclusion contrasts with that of a
previous study [L\'opez et al., Phys. Rev. E {\bf 80}, 040105 (R)
(2009)] which indicates that the transition at intermediate
density belongs to the $q=1$ Potts universality class. Almarza et
al. attributed the discrepancy to the use of the density as the
control parameter by L\'opez et al. The present work shows that
this suggestion is not sufficient, and that the discrepancy arises
solely from the use of different statistical ensembles. Finally,
the necessity of making corrections to the scaling functions in the
canonical ensemble is discussed.

\end{abstract}


\maketitle

The isotropic-nematic (IN) phase transition in a model of
self-assembled rigid rods (SARRs) on a square lattice was
considered for the first time by Tavares et al. \cite{Tavares}.
Using a theoretical approach and Monte Carlo (MC) simulation, the
existence of a continuous phase transition was pointed out.
However, the universality class of the transition was not studied
and the ordering of SARRs was assumed to be that of monodisperse
rigid rods (RRs), which was found to be the two-dimensional (2D) Ising class
\cite{Matoz}.

The criticality of the SARRs model in the square lattice was
investigated in Ref. \cite{Lopez} by means of canonical MC
simulation and finite-size scaling theory. The existence of a
continuous phase transition was confirmed. In addition, the
determination of the critical exponents along with the behavior of
the Binder cumulant ($g_4$) for different system sizes revealed
that the universality class of the IN transition, at intermediate
density, changes from 2D Ising-type for monodisperse RRs without
self-assembly to $q=1$ Potts-type (random percolation) for
polydisperse SARRs.

Recently, a multicanonical MC method based on a Wang-Landau
sampling scheme was used by Almarza et al. \cite{Almarza} to
reinvestigate the critical behavior of the model studied in Refs.
\cite{Tavares,Lopez}. Employing the crossing point of the Binder
cumulants ($g_4^*$) and the value of the critical exponent of the
correlation length ($\nu$), it was observed that the criticality
of the SARRs model is in the 2D Ising class, as in models of
monodisperse RRs. This finding is in sharp contrast to that
reported in \cite{Lopez}, and the authors have given a possible
explanation for this discrepancy \cite{Almarza} 
($\mu$ denotes the chemical potential): \emph{\textquotedblleft In the
analysis of L\'opez et al., the use of the density as the control
parameter leads to a value of the $g_4$ crossing that differs
substantially from that of the 2D Ising universality class. We
have shown that using $\mu$ as the control parameter leads to a
more robust scaling of $g_4$ and to a much better overall Ising
scaling.\textquotedblright}

The purpose of this Comment is to show that the above explanation
is insufficient, and to point out and discuss the source of
the discrepancy between our results and that obtained by Almarza et
al. As in Ref. \cite{Almarza}, the distinction between the two
universality classes is based on the determination of both the
value of $g_4^*$ and the value of $\nu$, which are clearly
different for the two universality classes under discussion.

Then, in order to analyze the explanation given by Almarza et al.,
a series of MC simulations have been conducted in the canonical
ensemble. The procedure was similar to that used in \cite{Lopez},
but this time maintaining as constant the surface coverage (at
$\theta=0.525$, critical density obtained in \cite{Lopez}) and
varying the temperature of the system (the natural control
parameter in the canonical ensemble).

The fourth-order Binder cumulant was computed as a function of the
temperature for different lattice sizes ($L \times L$), at
$\theta=0.525$ (see Fig. 1). The values obtained for the critical
temperature and the intersection point of the cumulants were 
$T_c=0.25$ and $g_4^*=0.639$, respectively. The same fixed value of
the cumulants was reported in \cite{Lopez}, which is consistent
with the $q=1$ Potts universality class (ordinary percolation). As
was mentioned in \cite{Lopez}, a value of $g_4^*\approx0.638$ was 
obtained by Vink for 2D site percolation
\cite{Vink}. Vink's result was recently reproduced by the
authors via Monte Carlo simulation \cite{foot}.

Once $T_c$ was calculated, the scaling behavior was tested by
plotting $g_4$ versus $\epsilon L^{1/\nu}$ (where $\epsilon$  is
the normalized scaling variable $\epsilon \equiv T/T_c - 1$) and
looking for data collapsing. Using the exact value of the critical
exponent of the correlation length for ordinary percolation, $\nu
=4/3$, an excellent scaling collapse was obtained, as shown in Fig.
1.

\begin{figure}[t]
\includegraphics[width=8cm,clip=true]{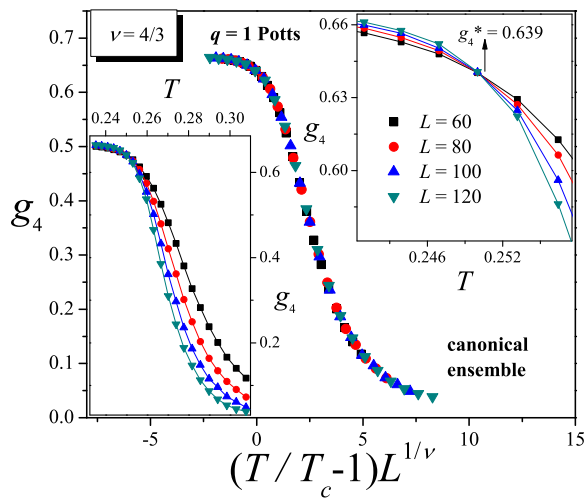}
\caption{(Color online) Data collapsing of the cumulant, $g_4$ vs $\epsilon L^{1/\nu}$.
Upper-right inset: Curves of $g_4(T)$ vs $T$ for lattices
of different sizes. From their intersections one obtained
$g_4^*$. In the lower-left inset, the data are plotted over a
wider range of densities. } \label{figure1}
\end{figure}

\begin{figure}[t]
\includegraphics[width=8cm,clip=true]{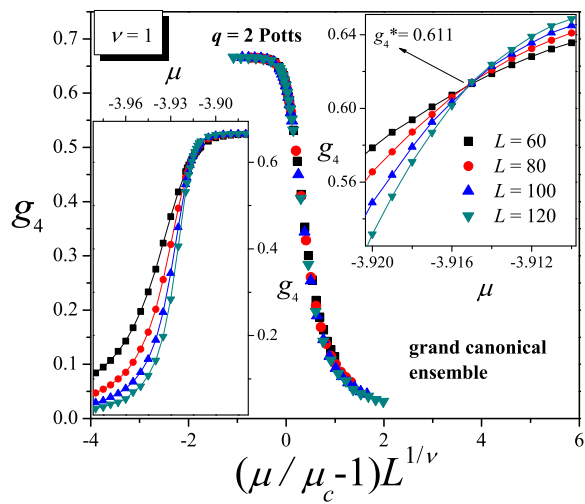}
\caption{(Color online) Data collapsing of the cumulant, $g_4$ vs $\epsilon
L^{1/\nu}$. Upper-right inset: Curves of $g_4(\mu)$ vs $\mu$ for
lattices of different sizes. From their intersections one obtained
$g_4^*$. In the lower-left inset, the data are plotted over a
wider range of densities. } \label{figure2}
\end{figure}

As mentioned above, the use of the density as the basic variable
in our previous study \cite{Lopez} led us to calculate values of
the critical exponents and the crossing point of the cumulants
that differ from those obtained by Almarza et al. \cite{Almarza}.
In Ref. \cite{Almarza}, the authors indicated that the difference
in the values of $\nu$ can be understood by introducing a
correction ($\ln L$) when using $\theta$ as the scaling variable.
However, the behavior of the curves in Fig. 1, where the control
parameter now is the temperature, provides convincing evidence
that the value of $g_4^*$ and the scaling obtained using $\nu =
4/3$ are not due to \emph{\textquotedblleft the use of the density
as the control parameter,\textquotedblright} as claimed in Ref.
\cite{Almarza}. This immediately suggests that the discrepancy
between the results of \cite{Lopez} and \cite{Almarza} arises from
the use of different ensembles. To test this new statement and, at
the same time, to check the data presented by Almarza et al., MC
simulations in the grand canonical ensemble were carried out using
an adsorption-desorption algorithm. It is important to note that
the algorithm used here is different from that used by Almarza et
al.

In the grand canonical ensemble, the critical behavior was studied
at the same point of the phase diagram, fixing the temperature at
$T=0.25$ and varying the chemical potential $\mu$. The Binder
cumulants versus $\mu$ are shown in Fig. 2. The intersection point
converges to a fixed point, allowing an accurate estimation of the
fixed value of the cumulants, $g_4^*=0.611$. This value is
consistent with the extremely precise transfer-matrix calculation
of $g_4^*=0.6106901(5)$ \cite{Kamieniarz} for the 2D Ising model.
On the other hand, very good collapse was obtained with $\nu=1$ in
the scaling plot of $g_4$ (Fig. 2), thus corroborating the data of
Almarza et al.

The results presented above confirm that the discrepancy under
study arises solely from the use of different statistical
ensembles. This behavior, which appears to be a violation of
the principle of ensemble equivalence, has been discussed many
times in the literature, usually related to systems subject to
constraint (such as the constraint of fixed density that is
imposed in canonical ensemble studies).

In this sense Fisher \cite{Fisher} (i) showed that, for systems
with thermodynamic constraints, critical exponents characterizing
scaling behavior at continuous phase transitions may deviate
significantly from their ideal theoretical counterparts (without
constraint) due to the effects of such constraints; and (ii)
established elegant relations between the exponents of the ideal
and constrained systems. In this scheme, known in the literature
as the \textquotedblleft standard Fisher renormalization
scheme,\textquotedblright the critical exponents in the
constrained system are renormalized if the specific-heat exponent
for the ideal system $\alpha$ is positive, or remain the same when
$\alpha$ is negative or zero.

The case presented here, where the system without constraint
belongs to the two-dimensional Ising universality class, shows
that a generalization of the Fisher renormalization is necessary
in certain circumstances where $\alpha=0$ (Dohm \cite{Dohm} has
also discussed this possibility for cases where $\alpha<0$).

In summary, several conclusions can be drawn from the present
Comment: (i) The discrepancy between the results in \cite{Lopez}
and \cite{Almarza} arises solely from the use of different
statistical ensembles. In this sense, even though it might be more
appropriate (or convenient) to use the grand canonical ensemble to
study a system such as the one described here, the consistency of the
results obtained in the canonical ensemble warrants an explanation
that has not yet been given. (ii) The system under study represents an 
interesting case where the use of different statistical ensembles
leads to different and well-established universality classes. (iii) 
Since most of the studies on the critical behavior of
self-assembled systems have been carried out in the canonical
ensemble, they should be revisited. (iv) Fisher
renormalization arguments predict that the values of the critical
exponents should remain unchanged since the specific-heat exponent
$\alpha$ for the present model is zero. However, our simulations
disagree with this prediction. The development of a modified (or
alternative) renormalization scheme, as was done in Refs.
\cite{Dohm,Ferreira}, could help to solve this problem. Obviously,
this task is beyond the scope of this Comment.

\acknowledgments This work was supported in part by CONICET
(Argentina) under project number PIP 112-200801-01332; Universidad
Nacional de San Luis (Argentina) under project 322000 and the
National Agency of Scientific and Technological Promotion
(Argentina) under project 33328 PICT 2005.




\begin{thebibliography}{99}
\bibitem{foot} In our simulations, we studied the site percolation problem on a square
lattice with periodic boundary conditions and used the
magnetization as the order parameter. As is standard for the
geometrical percolation problem, the magnetization was defined by
assigning the Ising spins on each cluster at random, with
probability $1/2$. Under these conditions, the cumulants touch at
the known critical threshold ($\approx 0.593$), giving
$g_4^*=0.636(4)$.
\end{thebibliography}
\end{document}